\documentclass[sigconf,nonacm]{acmart}
\AtBeginDocument{%
  }

\usepackage{textcomp}

\usepackage{paralist}     
\usepackage{tikz}
\usepackage{amsmath}
\usepackage{hyperref}
\usepackage{cleveref}
\usepackage{booktabs}
\usepackage[inline]{enumitem}
\usepackage{colortbl}
\usepackage{svg}
\usepackage{multirow}    
\usepackage{hyperref}
\usepackage{lscape}
\usepackage{tabularx}
\usepackage{footnote}
\usepackage[most]{tcolorbox}

\newenvironment{longdescription}
  {\begin{description}[style=unboxed]}
  {\end{description}}

\begin{document}

\title{A Systematic Review of Security Communication Strategies: \\Guidelines and Open Challenges}

\author{Carolina Carreira}
\affiliation{%
  \institution{Carnegie Mellon University, IST University of Lisbon and INESC-ID}
  \city{Lisbon}
  \country{Portugal}}
\email{carolinacarreira@cmu.edu}

\author{Alexandra Mendes}
\affiliation{%
  \institution{INESC TEC, Faculty of Engineering, University of Porto}
  \city{Porto}
  \country{Portugal}}
\email{alexandra.mendes@inesctec.pt}

\author{João F. Ferreira}
\affiliation{%
  \institution{INESC-ID and IST, University of Lisbon}
  \city{Lisbon}
  \country{Portugal}}
\email{jff@inesc-id.pt}

\author{Nicolas Christin}
\affiliation{%
  \institution{Carnegie Mellon University}
  \city{Pittsburgh}
  \state{Pennsylvania}
  \country{USA}}
\email{nicolasc@cmu.edu}

\renewcommand{\shortauthors}{Carreira et al.}

\begin{abstract}
 Cybersecurity incidents such as data breaches have become increasingly common, affecting millions of users and organizations worldwide. The complexity of cybersecurity threats challenges the effectiveness of existing security communication strategies. Through a systematic review of over 3,400 papers, we identify specific user difficulties including information overload, technical jargon comprehension, and balancing security awareness with comfort. Our findings reveal consistent communication paradoxes: users require technical details for credibility yet struggle with jargon and need risk awareness without experiencing anxiety. We propose seven evidence-based guidelines to improve security communication and identify critical research gaps including limited studies with older adults, children, and non-US populations, insufficient longitudinal research, and limited protocol sharing for reproducibility. Our guidelines emphasize user-centric communication adapted to cultural and demographic differences while ensuring security advice remains actionable. This work contributes to more effective security communication practices that enable users to recognize and respond to cybersecurity threats appropriately.
\end{abstract}

\ccsdesc[500]{Human-centered computing~HCI design and evaluation methods}
\ccsdesc[500]{Security and privacy~Human and societal aspects of security and privacy}
\keywords{Usable Security, Usability, User Study, Security, Privacy}

\received{20 February 2007}
\received[revised]{12 March 2009}
\received[accepted]{5 June 2009}

\maketitle

\section{Introduction}

With the rise of massive breaches and social engineering attacks, users must navigate an increasingly complex cybersecurity landscape, and cybersecurity has become a significant concern for individuals and organizations. For example, a 2023 leak in Twitter led to more than 235 million users' information being exposed~\cite{fung_2023}. However, these attacks are not limited to user data and can create losses of millions of dollars \,---\, for example, the Equifax data breach~\cite{srinivasan2019data}. Moreover, attackers often resort to social engineering tactics, exploiting users' inexperience to access sensitive information. For example, in March 2022, hackers compromised the Ronin Network, stealing approximately US\$620 million in cryptocurrency through a fake LinkedIn job offer~\cite{robertson_2022}. Although developers implement security safeguards to prevent data breaches, these measures are only effective when end-users understand and correctly respond to them. %

Effective communication strategies empower end-users to make informed decisions and reduce their vulnerability to threats. However, while security and privacy information is widely available to users, it can be challenging for many individuals to determine how to balance their privacy needs with other considerations.
In fact, most users feel uncertain when balancing their privacy. An obvious source of privacy uncertainty arises from incomplete security information~\cite{acquisti2015privacy}.
It is also important to motivate and educate users about security practices. Prior work has addressed this issue and studied security communication~\cite{bravo2014improving, Furnell_Jusoh_Katsabas_2006,kelley2009nutrition,Raja_Hawkey_Hsu_Wang_Beznosov_2011,Redmiles_Warford_Jayanti_Koneru_Kross_Morales_Stevens_Mazurek_2020,Schaub_Balebako_Cranor_2017,acquisti2015privacy}. Communicating with users can empower them and enable them to make better and more secure decisions~\cite{gates2013effective,zhou2018mobile}.

This review aims to identify effective strategies for communicating security concepts to end-users, bridging the gap between technical safeguards implemented by developers and the everyday decisions made by users.
Prior work that attempts to increase and improve communication on security topics for users can be roughly divided into two categories:

\begin{longdescription}
    \item[Papers that evaluate current communication strategies.]
    
        For example, Redmiles et al.~\cite{Redmiles_Warford_Jayanti_Koneru_Kross_Morales_Stevens_Mazurek_2020} conducted a user study focused on the quality of security and privacy advice on the web. Other work has tried to address this problem in other contexts, such as security warnings~\cite{bravo2014improving,Furnell_Jusoh_Katsabas_2006}.

    \item[Papers that suggest new ways to communicate.]
    
        Most work in the this category also includes suggestions for improving communication. A concrete example by Schaub et al.~\cite{Schaub_Balebako_Cranor_2017} analyzes why existing privacy notices fail to inform users and tend to leave them helpless. The authors also discuss principles for designing more effective privacy notices and controls.
        Other papers, however, like Kelley et al.~\cite{kelley2009nutrition}, suggest alternative communication methods. The authors suggest that security concepts may be displayed to users as a security label (similar to a nutrition, warning, or energy label). Raja et al.~\cite{Raja_Hawkey_Hsu_Wang_Beznosov_2011} studied another approach\,---\,they designed iterative firewall warnings in which the functionality of a personal firewall was visualized using a physical security metaphor. 

\end{longdescription}

However, there is a lot of overlap, and multiple studies survey similar populations (e.g., adults in the US) on similar topics (e.g., privacy policies) and with similar methodologies (e.g., surveys) as stated in other systemization of knowledge papers~\cite{distler2021systematic, wei2024sok}. \textbf{These papers are dispersed across different niches, like security warnings~\cite{zaaba2014study,aneke2021help,molyneaux2019understanding} and privacy policies~\cite{Carrion_Senor_Fernández-Alemán_Toval_2012,liao2020measuring,bracamonte2020effects}, but share common goals\,---\,how to improve security communication. } 
In this work, we investigate how security information is communicated to users. Our main contributions are:
\begin{compactitem}
    \item Producing the first systematic overview of how security communication is being studied across security domains;
    \item Deriving seven general guidelines to improve security communication across domains based on peer-reviewed literature; and
    \item Identifying a set of open problems on security communication that future work should address.
\end{compactitem}

We next present our scope and research questions section, reviewing existing literature and identifying gaps our research aims to address. In the method section, we describe our research design in detail, including the search string used, sources consulted, the criteria for including or excluding studies, the removal of duplicates, coding all papers, and developing a taxonomy. Afterward, we discuss our results and discussion, presenting the results of our analysis and highlighting key insights. Finally, we conclude the paper with a summary of our research, its implications, and potential avenues for future exploration.

\begin{figure*} 
\centering
\includegraphics[width=0.95\textwidth]{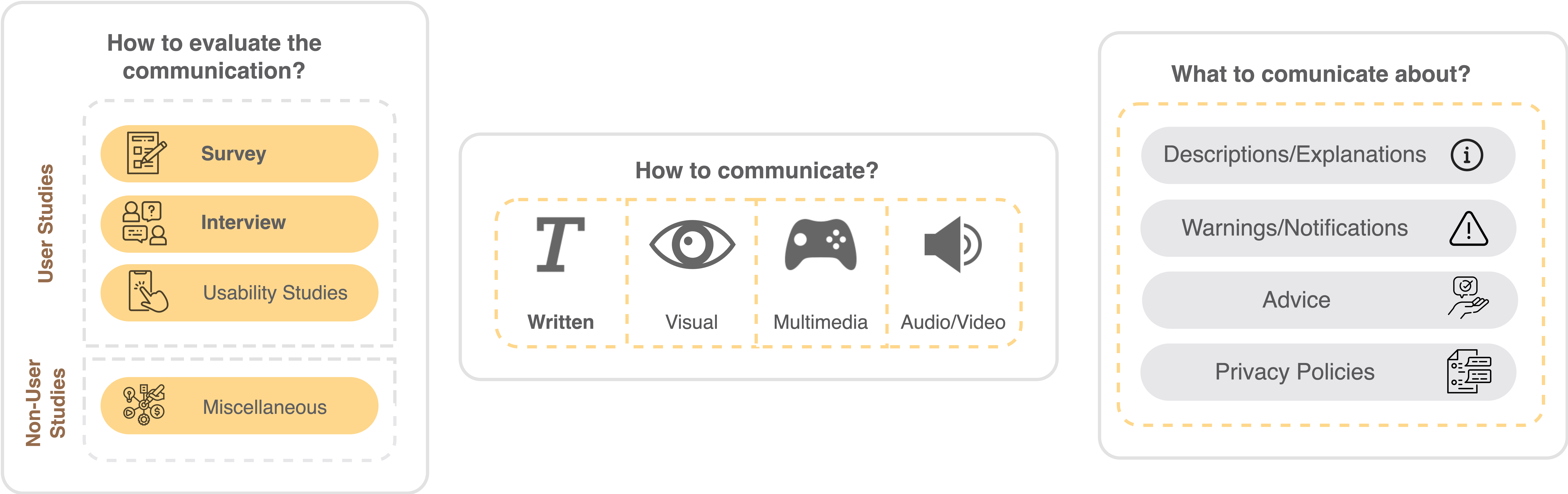}
\caption[]{Visual representation of the main research questions.}
\label{fig:rqs}
\end{figure*}

\section{Scope and Research Questions}

In this section, we describe our research goals and go over some related work. 

\subsection{Research Questions}

We address the following research questions.
\paragraph{\textbf{RQ1.  What key recommendations does the literature provide for improving security communication?}}
The main contribution of our paper is a set of self-contained and practical advice for improving security communication according to published research from the security community. Using a systematic taxonomy and analysis, we distill the advice for future work given by the papers we study. 
RQ1 helps clarify which strategies are most promising for improving security communication and for enabling security educators and developers to prioritize efforts where they will have the greatest user impact.

\paragraph{\textbf{RQ2. What techniques are used to study users' understanding of technical concepts?}} 

One of our primary objectives is to explore the approaches proposed in studies concerning the communication of security concepts. This is essential for recognizing patterns and gaps in the existing literature, which, in turn, helps inform the development of more effective communication strategies to meet users' needs and enhance their security awareness and behavior.
In this research question, we analyze various research methods used in these studies, including interviews, surveys, and case studies. We also examine the types of data collection and analysis techniques used in the literature, such as qualitative and quantitative approaches. We hope to understand how researchers tackle the challenge of communicating security.
Furthermore, we hope that by identifying the strengths and weaknesses of different approaches, we can help inform the development of more effective communication strategies that better meet users' needs.

\paragraph{\textbf{RQ3. Which communication techniques are used to communicate about security?}}

With this research question, we wish to understand the various communication techniques used in the literature, including textual communication, visual communication, labels, and other forms of communication. We also examine the types of media used to deliver security information, such as videos, graphics, and animations.

\paragraph{\textbf{RQ4. What are the security communication problems identified?}}
Our goal with this research question is to understand pressing security communication problems and highlight areas where further research is needed. 
By collecting the security communication problems that have been identified and those that have not yet been addressed, we can better understand the challenges practitioners face in communicating security information to users. 

\subsection{Empirical Studies}

Numerous empirical studies have examined the effectiveness of various types of security communication, such as warnings, alerts, and text messages. For example, one study found that text message alerts effectively increased individuals' compliance with emergency evacuation orders during a wildfire~\cite{kuligowski2018review}. %
Security communication is a crucial aspect of security systems as it is the primary means to notify users of potential security threats and breaches. The effectiveness of security communication impacts the success of security measures in mitigating security risks~\cite{carreira2021exploring, carreira2022studying}. As such, we argue that developing a comprehensive understanding of how security communication functions and how it can be improved is essential.

Previous studies explored various aspects of security communication, including its effectiveness, the types of messages used, and the factors that affect its success. One such study by Downs et al.~\cite{downs2007behavioral} investigated the impact of different warnings on users' behavior during a phishing attack. The study found that the perceived severity of the consequences does not predict behavior. They suggest that educational efforts should increase users' intuitive understanding rather than merely warning them about risks.
Another study by Furnell et al.~\cite{furnell2018enhancing} explored the effectiveness of various types of authentication messages in promoting secure behavior. The authors concluded that messages that emphasized the importance of security were effective.
Moreover, the users' cognitive and emotional states also influence the effectiveness of security communication. For example, a study by Van Boven and Loewenstein~\cite{knee2003implicit} found that more anxious individuals were more likely to take action to avoid a security threat.

Overall, the literature suggests that effective security communication involves using clear, concise messages emphasizing the threat's severity and the importance of secure behavior.  %

\subsection{Previous Surveys}

Previous work has attempted to partially analyze how communication is done on this topic. In this section, we address several systematic literature reviews. Hancock et al.~\cite{Hancock_Kaplan_MacArthur_Szalma_2020} did a 2020 meta-analysis on security warnings that aim to qualify their impact on users' behavior. Their work, however, was limited to security warnings. 
Lennartsson et al.~\cite{Lennartsson_Kävrestad_Nohlberg_2020} attempted to do a thematic literature review on the topic of usable security. The authors cite communication as particularly important in secure software. Their scope is broader than security communication. 
 In another domain, that of Password Managers, Chaudhary et al.~\cite{Chaudhary_Schafeitel-Tähtinen_Helenius_Berki_2019} did a systematic literature review to present suggestions for realizing a useable, secure, and trustworthy password manager. They argue that by bridging the communication/cognitive gaps between Password Managers designers and users, trust can be built between them~\cite{Chaudhary_Schafeitel-Tähtinen_Helenius_Berki_2019}. 
While related, the works mentioned in this section do not address our research goals with this literature review. These works are spread across domains (i.e., warnings~\cite{Hancock_Kaplan_MacArthur_Szalma_2020}, password management~\cite{Chaudhary_Schafeitel-Tähtinen_Helenius_Berki_2019}, and usable security~\cite{Lennartsson_Kävrestad_Nohlberg_2020}) which leads to a fragmented focus that does not address our cross-domain research questions.

While efforts have been made to improve communication, no systematic literature review explains security issues to a broader audience. 
We aim to synthesize a more cohesive understanding of security communication strategies by bridging these different areas.

\section{Method}

\begin{table}
\centering
\caption{The number of papers obtained from each database.} 
\label{tab:papers_databases}
\begin{tabular}{ll}
\toprule
\textbf{Database}                & \textbf{Papers}  \\ 
\midrule
Scopus   Science Direct & 1,186     \\
World Of   Science      & 1,424     \\
ACM   Digital Library   & 363    \\
IEEE   Xplore           & 512     \\ 
\midrule
\textbf{Total   Articles}        & \textbf{3,485}    \\
\bottomrule
\end{tabular}
\end{table}

We provide a detailed account of the transparent and replicable process that we followed in selecting and extracting data from the included studies. %

We describe a transparent and replicable process that we followed to identify and select relevant studies to answer our research questions. %

\subsection{Search String}

Our first step was to identify keywords that captured the essence of our research, after which we determined which digital libraries to search for publications.
Before deciding on the search string, we reviewed background work on security communication. This first ad-hoc literature review helped the authors better understand the subject.

We then focused on choosing the search string. We iterated over 12 different search strings. Each of these strings was taken into consideration by the team and compared. %
Our criteria for choosing the search string was that we wanted a search string that:
\begin{itemize}

    \item matched the most amount of relevant papers; %
    \item had keywords related to security communication;
    \item minimized the amount of non-relevant papers (e.g., if we added the word ``user'' to the search string, we could catch many papers that had nothing to do with security communication).
\end{itemize}

\begin{figure*} 
\centering
\includegraphics[width=0.66\textwidth]{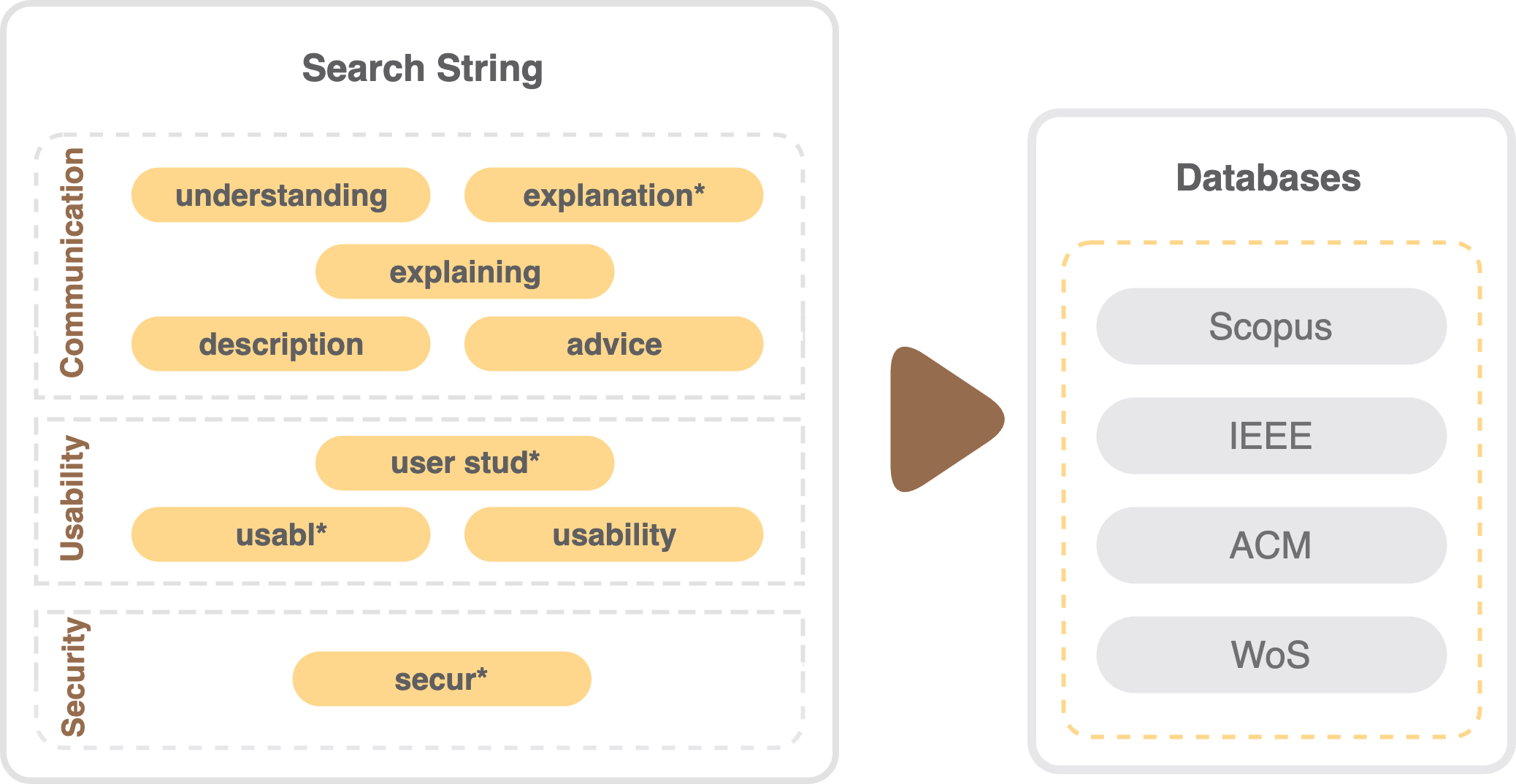}
\caption[]{Visual representation of the keywords included in the individual search strings.}
\label{fig:searcchstring}
\end{figure*}

To ensure that our search query was thorough, we used a conjunction (boolean ``AND'') of these three groups (i.e., the papers had to have a word from all three groups). Within each group, we used a disjunction (boolean ``OR'') of all the synonyms (e.g., for communication, we have ``communication'' or ``explanation''). Our keywords are divided into three groups: Usability, Communication, and Security.
We describe our search string in~\Cref{fig:searcchstring}.

We explicitly excluded the term ``communication'' from our search query. This decision was driven by the need to balance comprehensiveness with specificity, as including ``communication'' added retrieving an overwhelming number of papers (over 1,500 extra publications)  with only tangential relevance to security communication. Preliminary searches indicated that relevant studies were adequately captured through related keywords such as ``understanding,'' ``explanation,'' and ``usability,'' thereby maintaining the focus on security communication without introducing excessive noise.

\begin{tcolorbox}[
    enhanced jigsaw,
    sharp corners,
    boxrule=0.5pt, 
    colback=black!5!white,   
    boxrule=0pt, 
    frame empty
 ]
\textbf{\textit{Search Query:}} \textit{( understanding  OR  explanation*  OR  explaining  OR  description OR advice )  AND  ( secur* )  AND  ( usabl* OR user stud* OR usability)}
\end{tcolorbox}

\textbf{Note on Wildcard Usage:}
To enhance replicability and transparency, we utilized the wildcard character * in our search terms. This allows the search engine to include all variations of a root word, thereby broadening the scope of our literature search. For instance, ``\textit{secur*}'' captures terms such as ``security'', ``secure'', ``securing'', etc., ensuring that relevant studies using different terminologies are not inadvertently excluded. 

\subsection{Databases}

To conduct this survey, we searched for publications in the four primary databases for computer science literature, namely, Scopus Science Direct\footnote{\url{scopus.com/}}, Web Of Science\footnote{\url{webofscience.com/}}, ACM Digital Library\footnote{\url{dl.acm.org/}}, and IEEE Xplore\footnote{\url{ieeexplore.ieee.org/}}. Our search was based on relevant keywords and was conducted in the publication's \emph{Title}, \emph{Abstract}, and \emph{Author Keywords} fields. We searched each database independently, ensuring that our inclusion and exclusion criteria were consistently applied. The number of papers obtained from each database can be seen in~\Cref{tab:papers_databases}.
The search query was systematically executed across all four selected databases—Scopus Science Direct, Web Of Science, ACM Digital Library, and IEEE Xplore.

\subsection{Selection of Relevant Articles}
To identify duplicates and select the relevant papers, we used Rayyan\footnote{Rayyan is a collaborative systematic review software aimed at enabling a more efficient systematic review of papers. \url{https://www.rayyan.ai/}}. In total, from the 3,485 papers identified, 1,244 were duplicates.

\begin{figure*} 
\centering
\includegraphics[width=0.88\textwidth]{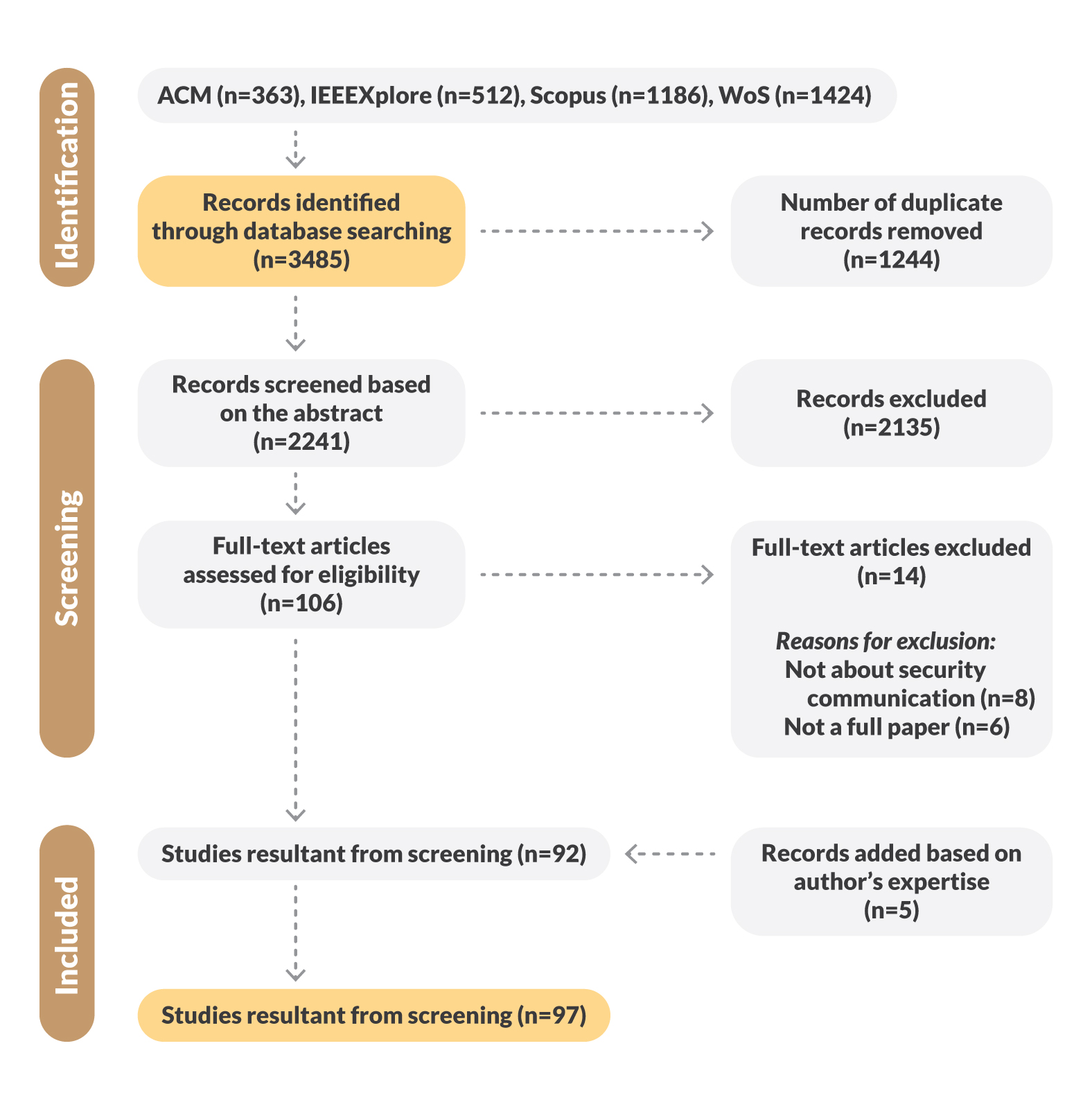}
\caption{Flow diagram for study selection process based on Moher et al.~\cite{prismaguidelines}}\label{fig:prisma}
\end{figure*}

Inclusion criteria included peer-reviewed articles, conference papers, and book chapters that focused on security communication to users, were published in English, and were available online. We excluded studies focusing on technical security aspects, such as cryptography or intrusion detection, and those unrelated to security communication (see~\Cref{tab:criteria}).

We applied the inclusion and exclusion criteria to all identified studies in two phases. In the first phase, two independent reviewers discussed the inclusion criteria and double-anonymized screened the titles and abstracts of the studies to identify potentially relevant articles using Rayyan. A third reviewer resolved disagreements between the first two reviewers. In total, the first two reviewers disagreed in 97 papers. After this review process, we eliminated 2,135 papers, and 106 remained.

In the second phase, one of the authors reviewed the full text of all the remaining 106 articles and excluded 14 according to the exclusion criteria previously mentioned (see~\Cref{tab:criteria} for more details). Finally, we included five articles that were not identified through our systematic approach but were pertinent to our research focus based on the author's knowledge. This is a common final step in surveys to augment the pool of papers aligning with the PRISMA methodology~\cite{Page_McKenzie_Bossuyt_Boutron_Hoffmann_Mulrow_Shamseer_Tetzlaff_Akl_Brennan_2021,prismaguidelines}. For a full breakdown of each review step see~\Cref{fig:prisma}. We finalized our selection of relevant articles with 97 publications, which can be consulted in~\Cref{app:fulllist}.

    \begin{table*}
\centering
\caption{Inclusion and exclusion criteria.} 
\label{tab:criteria}
\begin{tabular}{ll} 
\toprule
\textbf{Inclusion Criteria}                                          & \textbf{Exclusion Criteria}                              \\ 
\midrule
Papers   about security communication in general & Non archival papers \\
Papers   that explored new ways to communicate about security & Not-peer-reviewed papers  \\
Papers about security warnings and notifications & Papers   focused on ``explaining'' other topics \\
Papers about security communication with users &  Papers   not in English\\

Papers about privacy policies  & \\
 Papers about security communication with experts  & \\

\bottomrule
\end{tabular}
\end{table*}

\subsection{Data Extraction and Analysis}

After selecting the relevant papers, we extracted data using a standardized data extraction form. This form included information about the authors, publication year, research design, and our taxonomy, as described below. We group our topics into \textbf{security communication} (communication category, communication method, and advice for future communication efforts), \textbf{research method} (Type of Study and Artifacts), \textbf{communication problem} (the problem that the paper tries to solve and any open issues that remain).

\paragraph{\textbf{Communication Category}}
We aggregate papers in security communication into common themes such as warnings about cybersecurity threats, advice regarding cybersecurity practices, privacy policy communication, or explanations or descriptions of technical concepts. We categorize the papers by their content to learn about trends in research topics and identify over-explored and underexplored research areas.

\paragraph{\textbf{Communication Method}}
To understand the distribution of communication methods, we assess whether the paper introduces a new communication method, examines an existing one, or explores a combination of both. Within this category, we also categorize papers according to the process of communication used, such as visual, audio, game-based, video, or written formats and combinations thereof. 

\paragraph{\textbf{Type of Study}}
This involves detailing the methodology employed in the study, such as surveys, task-based assessments, interviews, or usability tests. It also includes the analysis approach, whether it involves statistical tests or coding methodologies, and which specific methods were used. We use this information to make comparisons across studies.

\paragraph{\textbf{Artifacts}}
Sharing artifacts promotes transparency, reproducibility, and future work. So, in this part of the taxonomy, we take note of the types of artifacts shared by the authors, such as interfaces or products tested, complete user study protocols, or fully anonymized user data.

\paragraph{\textbf{Problem that the paper tries to solve}}
This category captures the specific problems each paper addresses within cybersecurity and privacy. This comes mainly from the paper's motivation and communications goals.

\paragraph{\textbf{Advice for future communication efforts}}
This category is related to one of the main contributions of our papers: the compilation of peer-reviewed advice from various studies to guide effective practices and strategies for communication efforts. So, for this category, we summarize any recommendations or guidelines proposed by each paper for improving future communication efforts.

\paragraph{\textbf{Open problems that remain}}
This final category notes any unresolved issues or challenges identified in the papers that call for further investigation to identify where future research efforts should focus.

\section{Results}
In this section, we review some of the main insights from the papers in our corpus and address each of our research questions.

\begin{figure}[t]
  \centering
  
  \caption{Paper distribution by year of publication with linear trend line.}\label{fig:year}
  \resizebox{8.5cm}{!}{
  \includegraphics{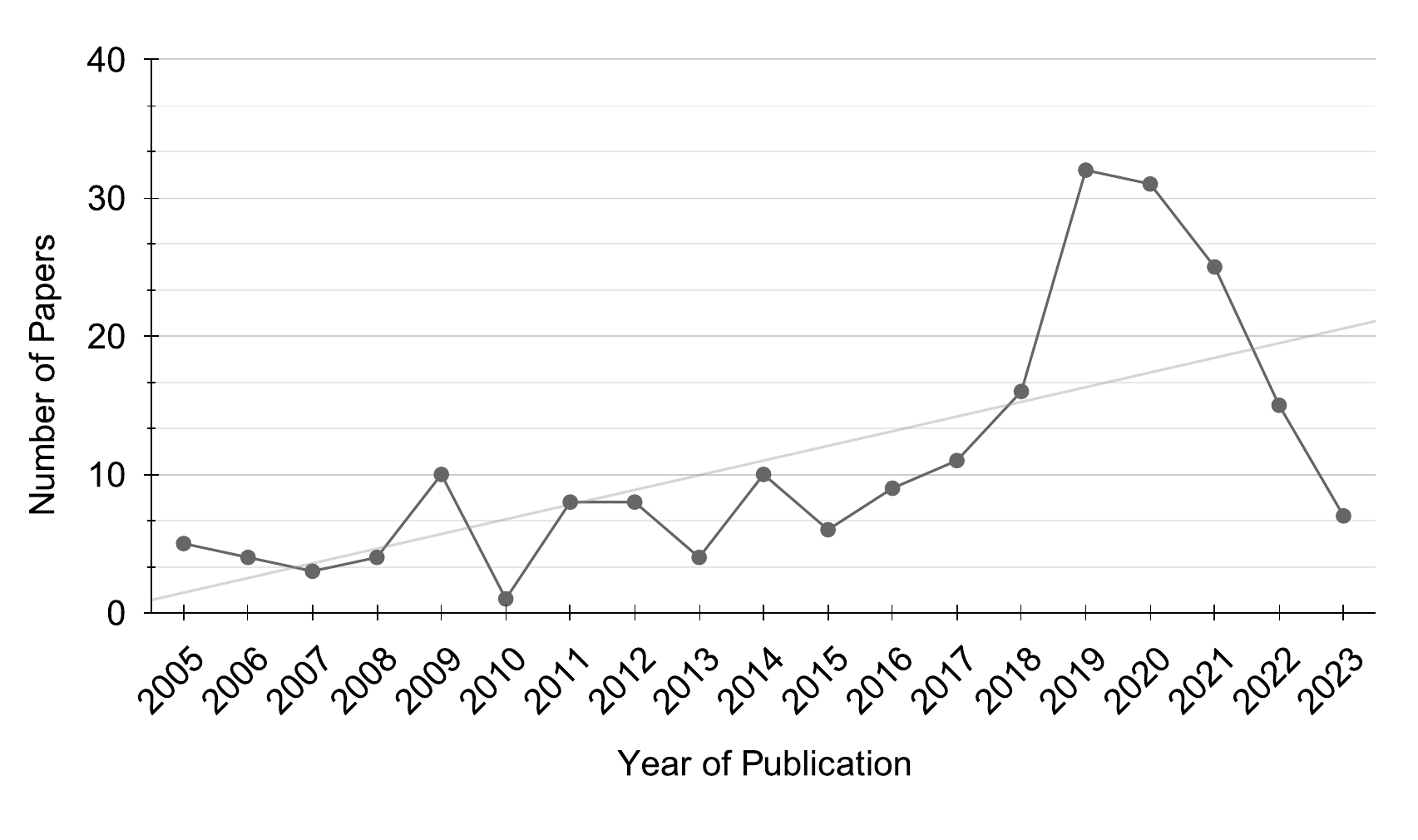}
}
  
\end{figure}
Our analysis of our corpus publication years reveals increased research output over time (from 2005 to 2021, see~\Cref{fig:year}). The data indicates that 2019 was the peak year of publication frequency (33\%, n = 32). This year was closely followed by 2020 (32\%, n = 31) and 2021 (26\%, n = 25). The years before 2019 saw fewer publications, with the median publication year being 2019. The oldest paper in our corpus is from 2005~\cite{hogganvik2005comprehension}, and the newest are from 2023~\cite{brunotte2023privacy,shrestha2023towards,zaaba2014study}.

\subsection{RQ1. Advice on Improving Security Communication}

In this section, we answer RQ1 by listing seven recommendations. 
The literature suggests that users have a better experience when they understand more about the technology~\cite{golla2018site}, 
so it is essential to increase user understanding of security concepts~\cite{kariryaa2021understanding,bracamonte2020effects}. 
Our analysis revealed that while many studies concur on the need for user-centric language and actionable guidance, subtle tensions emerged. Some papers suggest that practitioners should incorporate technical terminology for credibility, yet others warn against the confusion it can create. The following seven recommendations reconcile these differing viewpoints, offering a cohesive set of guidelines that practitioners can adapt to their unique contexts.
We distill advice for communicating security in our corpus under the following categories: Design and Presentation, Understanding, Personalization, and Behaviour.

\subsubsection{Design and Presentation}
Some papers~\cite{malik2019enhancing,wu2006security,yi2020appraisal,iacono2017signalling,von2022builds,bahrini2020enhancing} emphasize the need for \textbf{attention-grabbing and comprehensible visual design elements} in security communications. This includes using effective icons, color schemes, infographics, and visual metaphors~\cite{Raja_Hawkey_Hsu_Wang_Beznosov_2011,stover2021investigating} to make security warnings and permissions more noticeable and easier to understand. Particularly with warnings and notifications, users may disregard a message if it is not attention-grabbing. This happened in Sobey et al.'s~\cite {sobey2008exploring} work, where one of the security indicators was completely unnoticed by participants of their study, and as such, participants never got to see an informational pop-up. Due to the importance of the information, it can be useful to use more than one communication channel to get participants' attention, for example, using both email and more immediate notifications~\cite{golla2018site}, or even LEDs and haptic vibration~\cite{ma2017understanding}.

This is also the case with written communication where using bullet points and bold fonts is important~\cite{abu2020evaluating}.
Written text should be engaging to participants and attention-grabbing. The tendency to go into overly technical detail in writing is partially the reason why some literature tries to innovate with comics~\cite{zhang2016rolecomics,srikwan2008using,zhang2014stopcomic}, interactive activities~\cite{suh2022privacytooninterative}, games~\cite{kido2020sd,sheng2007anti,schufrin2022netvisgame,zargham2019could}, and even humor~\cite{zargham2019could} together with explanations.

\begin{tcolorbox}[
    enhanced jigsaw,
    sharp corners,
    boxrule=0.5pt, 
    colback=orange!9!white,   
    boxrule=0pt, 
    frame empty
 ]
\textbf{Recommendation 1:} Use attention-grabbing design elements to get users' attention.

\end{tcolorbox}

While most papers in our sample predominantly use written communication, \textbf{iconography} can complement text-based methods and is used in some papers in our corpus~\cite{bahrini2020enhancing,kelley2009nutrition,habib2021toggles,graham2021developing,yoo2019visual,kelley2009designing,stover2021investigating,ur2017design,zhang2014stopcomic}. 
Practitioners can use iconography to communicate with users quickly ~\cite{lazar2017research}. However, if communicators choose to use icons and symbols, these should be done with intention and care. Iconography or icons should be recognizable and understandable, reducing the cognitive load on users and facilitating quicker and more effective decision-making~\cite{samsudin2016symbolism}. 

It is crucial to acknowledge that icons are not universally applicable. While icons can enhance understanding, they do not replace all written content, especially with detailed and nuanced concepts. Ibdah et al.~\cite{ibdah2021should} found that some participants preferred to be informed through text over video. Moreover, interfaces incompatible with screen readers restrict non-sighted users’ ability to analyze information and make informed, secure decisions~\cite {napoli2018developing,Furnell_Jusoh_Katsabas_2006}. Where used, iconography should be consistently styled and tested with user groups to ensure they are effective and accessible.
Where researchers decide to communicate is also important. Deciding where to place information or icons is critical and can have an impact on users~\cite{petelka2019put}.

\begin{tcolorbox}[
    enhanced jigsaw,
    sharp corners,
    boxrule=0.5pt, 
    colback=orange!9!white,
    boxrule=0pt, 
    frame empty
 ]
\textbf{Recommendation 2:} Make sure any iconography is purposeful and understandable to users.
\end{tcolorbox}

The literature consistently advises using clear, straightforward language and avoiding unclear terminology~\cite{yi2020appraisal,bailey2021have,fassl2021exploring,napoli2018developing,samsudin2016symbolism,emang2019preliminary,desolda2023explanations,wu2020risk,greene2017must,abu2020evaluating}. Our corpus has examples of this, such as improving SSL certificate information in browser interfaces~\cite{biddle2009browser} and reorganizing permission categories to be more understandable~\cite{malik2019enhancing}.

Moreover, the communication should be clear in the sense that the communication channel should be straightforward~\cite{zargham2019could,liao2020measuring}. If you use text or iconography, you should communicate consistently across communication methods. If your system communicates through voice (e.g., a voice assistant~\cite{liao2020measuring}), it should consistently do so. 

Similarly, the wording chosen to describe security can positively and negatively influence users’ perceptions of security tools, thereby affecting the likelihood of adopting it~\cite{akgul2021secure}. For example, Redmiles et al. ~\cite{redmiles2016think} recommend that researchers avoid associations with marketing when explaining security. %
On the other hand, Distler et al.~\cite{distler2020making} suggest that using slightly technical vocabulary (e.g., ``encrypting'', ``securing'') felt reassuring and professional to participants. So, a balance must be reached between using some technical language and not overwhelming the user.

\begin{tcolorbox}[
    enhanced jigsaw,
    sharp corners,
    boxrule=0.5pt, 
    colback=orange!9!white,
    boxrule=0pt, 
    frame empty
 ]
\textbf{Recommendation 3:} Use clear and straightforward language while avoiding overtechnical language.
\end{tcolorbox}

\subsubsection{Understanding}

Users want to understand and want to be involved in the decisions regarding their technology~\cite{mahmoud2012does,railean2021onlite,balebako2013little}. Users are also interested in security topics and want to learn more about them~\cite{kariryaa2021understanding,brunotte2023privacy,railean2021onlite}. The literature suggests that even complex concepts can be explained~\cite{distler2022complex}, and if end-users cannot understand what they are disclosing or deciding about, they cannot be expected to use privacy mechanisms effectively~\cite{coopamootoo2011systematic}. So, researchers should not refrain from explaining security to users. 

The goals of communication should be not only to change behavior but also to mitigate misunderstandings about how security tools work (e.g., misunderstanding about the security of browser extensions~\cite{kariryaa2021understanding}). The system's complexity (e.g., smart voice assistants) can lead to inconsistent information~\cite{liao2020measuring}. The naming of specific features (e.g., private browsing mode~\cite{wu2018your}) can also induce users into a false sense of security.

Some papers use \textbf{alternative ways to communicate} security concepts, such as interactive games (e.g., Anti-Phishing Phil~\cite{sheng2007anti}) and multimedia approaches (comics combined with textual explanations~\cite{zhang2014stopcomic,srikwan2008using}), and applications~\cite{bahrini2019happypermi} to enhance learning outcomes and engage users more effectively. In some contexts, using statistics to talk to users about security\,---\,despite technical\,---\,can help improve users' understanding~\cite{kraus2014using}.
Regardless of the communication methods, incorporating explanatory information, such as why a password might be weak or what a specific permission entails, helps enhance user understanding and trust~\cite{abu2020evaluating,zhou2018mobile}. Providing reasons behind security advice, like explaining the benefits of enabling 2FA~\cite{golla2018site} or the risks of not updating software, helps make the advice more actionable and trustworthy.
Embedding security educational content within tools~\cite{zhou2018mobile}, such as password strength meters~\cite{xu2019explainable,ur2017design} or within the permissions granting process~\cite{zhou2018mobile}, can empower users by providing timely guidance during critical decision-making moments.

\begin{tcolorbox}[
    enhanced jigsaw,
    sharp corners,
    boxrule=0.5pt, 
    colback=orange!9!white,
    boxrule=0pt, 
    frame empty
 ]
\textbf{Recommendation 4:}  Users are curious and want to learn about security, so do not refrain from educating them.
\end{tcolorbox}

\subsubsection{Personalization}

Often, one explanation does not fit all. Several studies~\cite{wu2019catering,emang2019preliminary,felt2012android,al2020women,Raja_Hawkey_Hsu_Wang_Beznosov_2011,das2020smart,greene2017must,brunotte2023privacy,fagan2018follow} suggest personalizing security information to the user's concerns and knowledge. There is no ``universal user,'' so it is important to differentiate between users and adapt the communication style to the user's familiarity with the subject or their specific security needs~\cite{fassl2021exploring}.  Previous personal experiences and dispositions towards trust can impact users' security decisions~\cite{reeder2018experience}. Wu et al.~\cite{wu2019catering} proposed a way of designing participant-specific security explanations of Android Apps that cater to participants' individual mindsets. Moreover, some users are more sensitive to privacy issues than others, which should be considered when communicating with them~\cite{redmiles2016think}.
Busse et al.~\cite{busse2019replication} identified effective but unrealistic cybersecurity practices (e.g., adopting password managers of 2FA) that could be good in theory but that users do not adopt. This was also the case in other studies where users make unsafe choices~\cite{good2005stopping}.

Security communication should inform users about risks and motivate them to take protective actions~\cite{bravo2011improving,das2020smart}, but some types of communication work better for certain types of users. When talking with them, it is essential to consider users' predispositions and knowledge. For example, for some users, amplifying stories of others’ negative experiences can be a good way to communicate about dangerous behaviors~\cite{redmiles2016learned,redmiles2016think}. And for others, familiarity with technology influences their behavior~\cite{kani2012increasing}.

So communicators should explain security to participants, and a suggestion from the literature is to use participatory design~\cite{hagen2009human}. Participatory design consists of improving communication prototypes with the users' iterative feedback~\cite{fassl2021exploring}.

\begin{tcolorbox}[
    enhanced jigsaw,
    sharp corners,
    boxrule=0.5pt, 
    colback=orange!9!white,
    boxrule=0pt, 
    frame empty
 ]
\textbf{Recommendation 5:} Personalize security information to the user (e.g., expertise level and specific concerns). 
\end{tcolorbox}

Our corpus recommends understanding the context in which users are making security decisions~\cite{kariryaa2021understanding,amran2017usable,bahrini2019happypermi,fagan2018follow,emang2020usable,gorski2018developers,al2021look,kelley2009designing,suh2022privacytooninterative}. Warnings, advice, and explanations should consider the user's environment and current activities.
Gorski et al.'s~\cite{gorski2021just} eye-tracking study demonstrates that the placement of security-relevant information in non-security API documentation significantly impacts developers' ability to find and use this information. Their research shows that security content placed in proximity to functional code examples is more readily discovered, as developers primarily focus on code snippets rather than prose when solving programming tasks. When developers are focused on completing their primary functional task, security considerations often become secondary unless deliberately integrated into their workflow.  As in other fields, when participants have a task in mind, they work to complete it, and security is not a priority~\cite{gorski2021just,wu2006security,napoli2018developing}.

Indeed, from an outside point of view, the decision to perform some action is binary; the user either does the action or does not. However, trust and behavior are not binary from the user's perspective. Ukrop et al.~\cite{ukrop2020will} directly identify this point. They studied developers' trust in flawed TLS certificates and found that the decision to trust, despite being binary, was actually informed by non-binary factors such as the timing of the warning. As such, it is important to personalize security communication to the context where it is being communicated~\cite{shrestha2023towards,gorski2018developers}.

Understanding the \textbf{context and reasons} behind the users' actions is important when designing future security communication. Communication with users about their reasons can address specific issues they face while avoiding habituation~\cite{shrestha2023towards,wu2006security,samsudin2016symbolism}. As such, before explaining a concept, researchers should first study the user's context\,---\,their location (office, home), goals (completing a task or just exploring), and the specific technology they are interacting with.
Reputation and previously established trust with websites or applications can make users make dangerous decisions even when faced with warnings~\cite{reeder2018experience}. 

Timing of security communications is another critical and often-mentioned issue. Determining the most effective moment to deliver security messages\,---\,whether before, during, or after user interaction with a system\,---\,remains a challenge, is crucial for the efficacy of these messages, and depends on the context~\cite{wu2006security,ukrop2020will,felt2012android}.

\begin{tcolorbox}[
    enhanced jigsaw,
    sharp corners,
    boxrule=0.5pt, 
    colback=orange!9!white,
    boxrule=0pt, 
    frame empty
 ]
\textbf{Recommendation 6:} Personalize security information to the user's context (e.g., what action is the user trying to accomplish and what is their environment like). 
\end{tcolorbox}

\subsubsection{Changing behavior}

Users also struggle to \textbf{change their behavior}. As such, it is easier to support an existing behavior over inducing change~\cite {fagan2016they}.
The literature advises that explaining the reasons for following security advice and giving actional advice are good practice~\cite{althobaiti2018faheem} to convince users to change practices. 
Educating users about how security works may increase motivation to practice secure behavior because it helps to justify the need~\cite{zhang2014stopcomic,gates2013effective,zhou2018mobile,redmiles2016think,abu2020evaluating,golla2018site,raja2009revealing,fagan2016they}. An example is password strength meters, these explaining why the passwords are weak or strong and work to change users behavior~\cite{xu2019explainable,ur2017design}.
Similarly, according to Zhou et al. ~\cite{zhou2018mobile} even a brief and informal security education can be effective and cost-efficient in providing the desired education to mobile app users. In their study, a significant percentage of study participants chose to use a stronger security measure after being educated on security.

Advice and communication that has the purpose of changing behavior should be \textbf{actionable} and not abstract~\cite{Redmiles_Warford_Jayanti_Koneru_Kross_Morales_Stevens_Mazurek_2020,hogganvik2005comprehension}. As such, if communicators want users to know something then they should directly mention it with clear language, be very explicit about what they want the users to know~\cite{desolda2023explanations,golla2018site}, and not refrain from explaining important concepts~\cite{distler2022complex,fulton2019effect}. %
In a study about security advice, Redmiles et al. ~\cite{redmiles2016learned} found that nearly 50\%  of users accept advice because they trusted the source. So, the source must be identified and demonstrably authoritative, such as via professional credentials.
Trusting something they should not trust can have severe consequences for users~\cite{ukrop2020will}. For that reason, ethical considerations should be addressed when trying to change users' behavior, as research on this topic can be used for multiple purposes~\cite{distler2020making}, including convincing users to trust untrustworthy advice.

\begin{tcolorbox}[
    enhanced jigsaw,
    sharp corners,
    boxrule=0.5pt, 
    colback=orange!9!white,
    boxrule=0pt, 
    frame empty
 ]
\textbf{Recommendation 7:} Have clear communication goals when talking with users. Security communication can change behaviors, inform, or be used to manipulate users.
\end{tcolorbox}

\subsection{RQ2. Type of Study and Techniques For Users' Understanding}\label{subsec:typeofstudy}

We categorize studies into two broad types: user studies and non-user studies. Most reviewed papers (92\%, n = 89) are user studies involving direct participant engagement to gather data on cybersecurity communication strategies. \textbf{Of the 89 user studies in our corpus, only about a fourth (24\%, n = 21) share their user study protocol}\,---\, e.g., their interview or survey script.  Non-user studies are less frequent and include case studies~\cite{garcia2009personalized}, sentiment analyses~\cite{aneke2021help}, evaluations of security tools without direct user interaction~\cite{shin2013supporting,napoli2018developing,coopamootoo2011systematic} or are mainly based on related work~\cite{zaaba2014study,srikwan2008using}.

The user studies also show a wide range in the number of participants involved, from as few as 4~\cite{reynolds2021user} to as many as 6,000~\cite{reeder2018experience}, with a median of 65 participants per study. The most common participant count was 60, observed in four separate studies. Studies with larger participant pools often employed quantitative analytical methods, while those with fewer participants preferred qualitative approaches. 
Despite their inherent subjectivity, qualitative evaluations are necessary to understand participants~\cite{fassl2021exploring,lazar2017research}.

The methodologies used in user studies in our corpus are diverse. Surveys are the most common method (74\%, n = 72). This may be linked to larger participant pools and a more quantitative approach. Interviews, on the other hand, were conducted in 30 studies. This research method is commonly used for deeper insights through direct interaction. For example, Abu et al.~\cite{abu2020evaluating} used interviews to understand users' mental models of private browsing. 

Another subsection of our corpus (29\%, n = 28) used task-based studies where participants interacted with a product or tool to perform specific tasks, facilitating observation of user behavior in controlled scenarios~\cite{sheng2007anti,paspatis2020appaware}. These were often used together with usability studies (14\%, n = 14) explicitly focused on the usability aspects of tools or systems. Similarly, many studies have combined these methods, such as tasks, surveys, and usability-oriented methodologies. A specific example is Iacono et al.'s~\cite{iacono2017signalling} work on notification for signaling over-privileged permissions. In this study, participants performed a task where they had to choose an application to download, interact with the prototype notification, and participate in a structured interview. 

A substantial number of studies that used qualitative methods opted for open, emergent coding to derive themes and insights. Qualitative studies commonly reported using double-anonymized coding procedures. On the other hand, quantitative studies used standard statistical tests such as ANOVA, Kruskal-Wallis, Mann-Whitney U tests, and various regression models, catering to both parametric and non-parametric data sets.

\begin{tcolorbox}[
    enhanced jigsaw,
    sharp corners,
    boxrule=0.5pt, 
    colback=black!5!white,   
    boxrule=0pt, 
    frame empty
 ]
\textbf{Answer to RQ2:} 
Our corpus combines surveys, interviews, and task-based studies. Each method brings unique strengths: surveys provide statistical strength by usually relying on a large sample, interviews uncover deeper issues, and task-based studies reveal practical challenges. 
\end{tcolorbox}

\subsection{RQ3. Communication Techniques}

\subsubsection{\textbf{Category}}
The most frequent type of security communication was \textbf{descriptions/explanations}, with 56 (57\%) papers. Papers in this category typically aim to assess how effectively different methods convey security information to enhance user comprehension and security behavior. 
The second most popular category was \textbf{warnings/notifications} (21\%, n = 20), and it typically explores their effectiveness. Warnings/Notifications are security communication that occurs when a human uses a product. This communication interrupts the regular use of a product to notify the user about a specific security issue. 
With eight papers, the \textbf{security advice} category addressed how users receive and respond to cybersecurity advice, including their trusted sources and the most persuasive formats. Seven of these papers focus on Privacy Policy. This category's relatively low number of papers suggests a potential research opportunity. %
Among the topics covered in our review, two papers specifically focused on the design and efficacy of password meters~\cite{xu2019explainable, ur2017design}. Password meters provide real-time feedback on the strength of user-selected passwords, aiming to encourage more robust, more secure password creation. These studies examine aspects such as the design elements of the meters (e.g., color-coded strength indicators, text feedback) and the overall effectiveness of these meters in influencing password complexity.
Finally, papers classified under \textbf{other} include topics that do not neatly fit the above categories or that cover broader conceptual studies, such as mental models~\cite{raja2009revealing}.

\begin{table}[]
\caption{Overview of the communication categories present in our corpus.}\label{tab:categories}
\centering
\begin{tabular}{@{}lll@{}}
\toprule
\textbf{Category}                  & \textbf{\begin{tabular}[c]{@{}l@{}}Number  \\of studies\end{tabular}} & \textbf{\begin{tabular}[c]{@{}l@{}}Percentage \\ of studies\end{tabular}} \\ \midrule
\textbf{Descriptions/Explanations} & 56                                                                              & 57.7\%                                                                          \\
\textbf{Warnings/Notifications}    & 20                                                                              & 20.6\%                                                                          \\
\textbf{Advice}                    & 8                                                                               & 8.2\%                                                                           \\
\textbf{Privacy Policy}            & 7                                                                               & 7.2\%                                                                           \\
\textbf{Other communication}       & 4                                                                               & 4.1\%                                                                           \\
\textbf{Password Meter}            & 2                                                                               & 2.1\%                                                                           \\ \bottomrule
\end{tabular}%

\end{table}

\subsubsection{\textbf{Communication Method}}
Our review distinguishes between papers proposing new communication methods and evaluating existing ones. A significant portion of papers (65\%, n = 63) introduce new ways to communicate cybersecurity concepts, suggesting interest in developing security communication strategies. On the other hand, 50 (52\%) papers focus on analyzing existing communication techniques, with 16 (17\%) of these papers evaluating current methods and proposing new ones. 
Our review also categorizes the papers based on the communication domains used to convey security concepts. The most prevalent communication domain was written communication (used in 85 papers). Textual communication is the traditional way to communicate complex cybersecurity issues like privacy policies, and it is also the one more prevalent in our corpus. The second most common method was visual communication, used in 52 papers, which often complements written content with icons, colors, and other visual aids to enhance understanding and retention. Many studies (46\%, n = 45) combine written and visual communication. 
We also found in our corpus some less common methods, like games~\cite{scholefield2019gamification} and comics~\cite{zhang2016rolecomics,srikwan2008using,zhang2014stopcomic} and even a smart keyboard~\cite{de2011does}, which explore more alternative ways to educate about cybersecurity. These methods often integrate multiple communication domains, including visual and textual elements.
Audio and video were the least represented forms of communication in our corpus, with only three papers using audio and two using video. 

\begin{figure*} 
\centering
\includegraphics[width=0.66\textwidth]{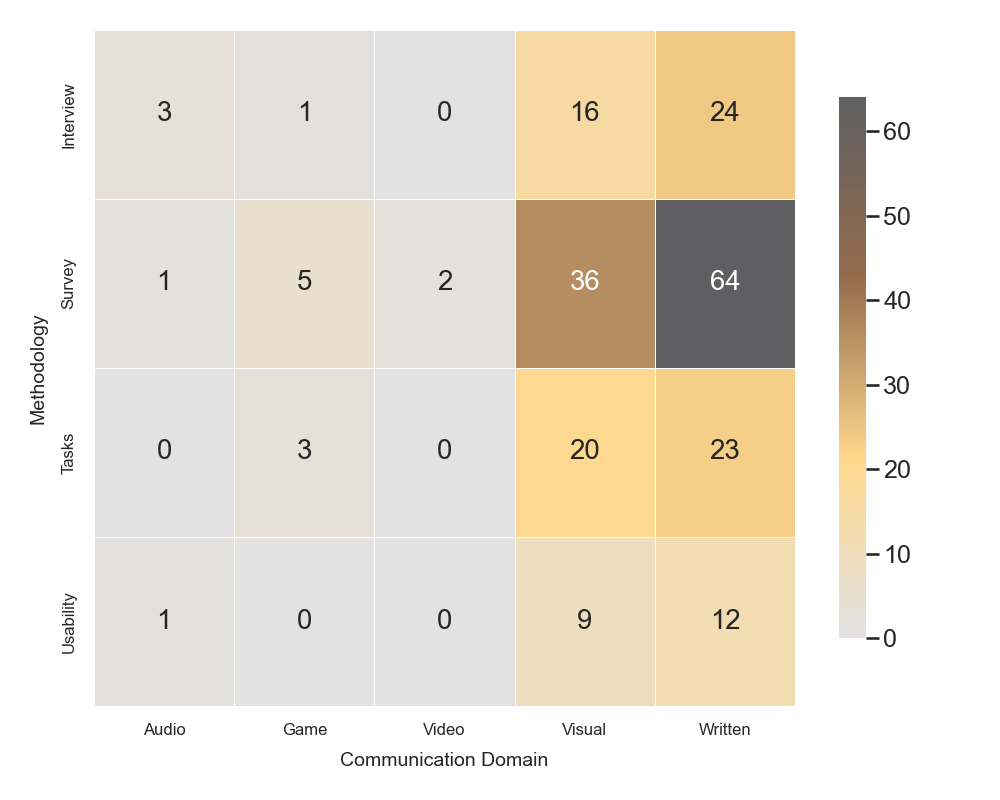}
\caption[]{Heatmap of the cross-section between Communication Domain and Methodology.}
\label{fig:heatmap}
\end{figure*}

Examining the intersection of methodologies (surveys, interviews, tasks, and usability testing) in~\Cref{fig:heatmap}, we can see that the majority of studies combine written and visual communication methods. This pattern persists across all methodological approaches.
Alternative communication methods such as games, audio, and video show minimal representation across all methodologies. This distribution substantiates our finding that security communication research primarily relies on traditional approaches, while alternative communication domains remain underexplored. The absence of significant research using audio, video, and game-based approaches presents opportunities for future diversification of security communication strategies.

\begin{tcolorbox}[
    enhanced jigsaw,
    sharp corners,
    boxrule=0.5pt, 
    colback=black!5!white,   
    boxrule=0pt, 
    frame empty
 ]
\textbf{Answer to RQ3:  } 
Most studies use descriptions and explanations to convey security concepts and primarily rely on written explanations\,---\, as opposed to 
other media (audio, video, visual).
\end{tcolorbox}

\subsection{RQ4. Communication Problems}

In this section, we answer RQ4 by going over communication problems identified in our corpus.
The main problem, or limitation, identified in our corpus was their study population. 
Some papers mention that a field study or a \textbf{larger participant pool} should be used to test the effectiveness of their communication method~\cite{zhou2018mobile,balebako2013little,ur2017design,de2011does,abu2020evaluating,von2022builds,reynolds2021user,gorski2021just,samsudin2016symbolism,kraus2014using}. Other papers argue that their suggestions for communicating can be improved and iterated in future work~\cite{tucker2015privacy,zhang2015towards,suh2022privacytooninterative,shin2013supporting,tan2014effect}. Increasing the participant pool can make studies gain statistical power and strengthen their results. 
Moreover, some of our corpus studies the same problems (e.g., privacy policies~\cite{paspatis2020appaware,garcia2009personalized}) and methods overlap. So, replication and iteration over previous work could prove valuable for the security communication domain. %

\begin{tcolorbox}[
    enhanced jigsaw,
    sharp corners,
    boxrule=0.5pt, 
    colback=red!5!white,   
    boxrule=0pt, 
    frame empty
 ]
\textbf{Future work insight:} Our corpus suggests that future work should focus on replicating existing literature (for example, with large populations) or iterating over existing work.
\end{tcolorbox}

On the other hand, in our corpus, we only found 
one replication paper~\cite{busse2019replication}. Reproducibility is extremely important in science, and other scientific fields, e.g., 
psychology~\cite{open2015estimating} or HCI~\cite{hornbaek2014once}, 
have struggled with replicating results.
Furthermore, as stated in~\Cref{subsec:typeofstudy}, only 24\% of the user studies in our corpus shared their protocols. This poses a challenge for future replication efforts. Thus, providing more artifacts to inform future replication efforts should be a priority.

\begin{tcolorbox}[
    enhanced jigsaw,
    sharp corners,
    boxrule=0.5pt, 
    colback=red!5!white,   
    boxrule=0pt, 
    frame empty
 ]
\textbf{Future work insight:} With only 24\% of studies sharing their protocols, our findings suggest that, to foster reproducibility, authors should share their artifacts and user studies protocols. 
This may require incentivization\,---\,e.g., 
in the form of awards or recognition for ease of 
reproducibility.
\end{tcolorbox}

The impact of local norms and cultural contexts on the adoption of security advice is an issue that warrants further investigation. Cultural factors such as attitudes toward authority, risk perception, and privacy can significantly influence how individuals interpret and act on security recommendations. Studies suggest understanding and incorporating local cultural elements into security advice could significantly enhance its effectiveness and acceptance ~\cite{al2020women}. However, most of the works in our corpus focus on the US and English-speaking countries ~\cite{shrestha2023towards,al2021look} and directly mention this as a limitation. 
Moreover, alternative communication, for example, through comics, must consider directional reading habits, such as left-to-right or right-to-left, which vary by language and culture ~\cite{zhang2016rolecomics}.  This geographical and cultural bias poses a risk of adopting a one-size-fits-all approach in cybersecurity communication. In regions where social norms, digital literacy, and trust in digital systems vary widely, security advice that is not culturally tailored may fail to resonate.

And there may be more specificities we have not yet identified. All of the papers in our corpus also focus on the general population and do not address \textbf{older adults or children}. These groups of users have increasingly used technology and have specificities that need to be addressed. Older adults are a particularly vulnerable population that has been understudied~\cite{frik2019privacy}. 

Moreover, the majority of the papers in our corpus use written communication. This seems to be a significant trend in cybersecurity communication. While written communication effectively conveys detailed and complex information, it also presents potential limitations. Relying heavily on written content may inadvertently exclude or disadvantage users with low literacy levels or those who are visually impaired~\cite{lazar2017research}. Such users might find it challenging to engage with security instructions or warnings that are predominantly text-based \cite{napoli2018developing}.
It is also important to recognize that our current understanding of these issues is limited by the lack of detailed demographic data in many studies. This gap makes it difficult to precisely quantify which populations are understudied. 

\begin{tcolorbox}[
    enhanced jigsaw,
    sharp corners,
    boxrule=0.5pt, 
    colback=red!5!white,   
    boxrule=0pt, 
    frame empty
 ]
\textbf{Future work insight:} 
Future work should be more inclusive and extend to more diverse populations (with a focus on a broader range of nationalities and cultures, age groups, and accessibility needs).
\end{tcolorbox}

The overwhelming majority of our corpus focuses on the short term with single-time studies and does not look at security communication longitudinally\,---\,only one study by Weinshel et al.~\cite{weinshel2019oh} conducts a longitudinal user study. 
Short-term studies may miss changes in behavior over time due to new threats or more user education. They might also capture immediate reactions or learning outcomes but fail to assess how well users retain and apply security knowledge over time. 
Moreover, people often get informed in various ways (e.g., school, friends, media) that are hard to analyze as a whole in a short-term study and thus are not analyzed in our corpus. 
The big challenge of longitudinal studies is that they are usually very time-intensive and expensive for researchers. However, some of our corpus directly identified the lack of long-term analysis as a shortcome~\cite {al2021look,zhang2016rolecomics}.

\begin{tcolorbox}[
    enhanced jigsaw,
    sharp corners,
    boxrule=0.5pt, 
    colback=red!5!white,   
    boxrule=0pt, 
    frame empty
 ]
\textbf{Future work insight:} 
Future work should strive to conduct longer-term studies to get longitudinal insights on security communication efforts.
\end{tcolorbox}

\section{Discussion}

Communicating effectively and efficiently about security remains an open challenge. In this section, we identify two communication paradoxes in security communication\,---\,the \textbf{Comprehension -- Jargon} and the \textbf{Awareness -- Discomfort} paradox and an additional two challenges that can be derived from our corpus\,---\,\textbf{Information Overload Challenge} and \textbf{Innovation Standardization Challenge}. While our systematic review identifies practical communication challenges, these can be better understood through established theoretical frameworks. In this section, we also connect our findings to relevant theories in communication science, psychology, and human-computer interaction to provide a stronger foundation for future research.

\subsection{Communication Paradoxes}
Striking the right communication \textbf{balance} is a big open problem in our corpus, but some of the insights may seem contradictory or paradoxical. 
During our synthesis of the literature, we noted recurring tensions in user communication: while some studies recommended technical detail for credibility, others showed such detail caused confusion. Similarly, while raising awareness of threats increased caution, it sometimes led to increased anxiety and distrust. We conceptualized these tensions as paradoxes to highlight their recurring nature and the need for careful balance.
We reason about them in this section.

\subsubsection{Comprehension -- Jargon Paradox}
 Users seem to want to understand and learn more about technology~\cite{balebako2013little}, and some papers suggest that using technical language is beneficial~\cite{kraus2014using,distler2020making}. However, overly technical jargon can also hinder understanding and trust in security mechanisms. As such, we argue that researchers should optimize their communication to strike a balance between going into the technical details of security and simultaneously not using too much jargon. 
 A balanced approach, such as using analogies or simplified definitions for complex terms, provides credibility without overwhelming users. Drawing on cognitive load theory~\cite{plass2010cognitive}, reducing extraneous complexity in language supports better user comprehension while maintaining a sense of security.

\paragraph{Mental Model Theory}
We can understand the Comprehension-Jargon paradox through mental model theory~\cite{mentalmodel-johnson}. In this theory, users construct simplified internal representations of complex systems to guide their decision-making. Cognitive dissonance occurs when security communications introduce technical jargon that conflicts with existing mental models. This explains why technically precise security information can paradoxically reduce comprehension.
Research by Wash~\cite{mentalmodel-wash} on folk models of security threats demonstrates how users develop simplified, often incorrect mental models of security mechanisms. It achieves higher effectiveness when communication aligns with these existing models while gradually correcting misconceptions. This suggests that security communications should elicit users' existing mental models before attempting to refine them.
    
\subsubsection{Awareness -- Discomfort Paradox}
Similarly, users want to understand the risks and benefits of security technology~\cite{mahmoud2012does,railean2021onlite,balebako2013little}, which can help them make more informed choices. However, explaining too much or too little can make them feel unsafe. For example, describing in detail how encryption works may make users feel unsafe~\cite{distler2020making}. Explaining too little is also not useful as it does not effectively inform users. There must be a balance between raising participants' awareness levels and thus creating a greater perception of risk, \textit{versus} reassuring users and making them feel more secure~\cite{fassl2021exploring}. We argue that finding the right balance when explaining security risks without overwhelming or frightening users is crucial for effectively understanding the system. Consequently, a \textit{dual} trust is gained: in the system and as users of the system. Relying on analogies with familiar, real-world situations, albeit imperfect, might be helpful in striking this balance. 
From a psychological perspective, this paradox can be viewed through the lens of risk communication. Effective risk communication strikes a balance between informing users about potential harms (to prompt action) and not inducing unnecessary stress. We argue that the information depth should be tailored and should use reassuring language or positive reinforcement to maintain trust while ensuring users remain vigilant.

\paragraph{Risk Communication Theory}
The Awareness-Discomfort paradox aligns with fundamental principles from risk communication theory. Fischhoff's~\cite{riskcommunication-fischhoff} framework for effective risk communication says that communication must balance informing users of potential threats while at the same time providing actionable paths to mitigation. This explains why security communications that present threats without clear remediation options often generate anxiety rather than action.
Sandman's~\cite {riskcommunication-sandman} risk perception equation (Risk = Hazard + Outrage) also corroborates this paradox. Security communications that emphasize technical hazards without addressing user concerns (outrage factors) fail to appropriately calibrate risk perception. This theoretical lens suggests that effective security communication should explicitly address both the objective security threat and the subjective concerns users experience when facing uncertainty.

\subsection{Additional Challenges}

In addition to these paradoxes, we also identified some future work dimensions that need to be balanced to improve effective communication.

\subsubsection{Information Overload Challenge}
Communicating too often to users and thus bombarding them with information can irritate them, negatively influencing their perception of security and usability. Too much communication can backfire and fatigue users~\cite{wu2020effects,felt2012android,wu2006security}. Not communicating at all is not an option, so here, too, a balance needs to be struck. A problem we found is that the majority of our corpus (65\%) suggests new ways to communicate, thus suggesting communicating more with users. However, our corpus also warns about information fatigue. As such, we can conclude that if all of these explanations and advice were to be implemented, they could overwhelm a user. Thus, a balance must be struck between increasing user education and not overwhelming users.

\subsubsection{Innovation Guidelines Challenge}

As mentioned before, most of our corpus used textual communication, which has challenges. 
Our review indicates that video and audio-based communication strategies are scarce in security communication research. 
Several factors may contribute to this trend. Firstly, producing high-quality multimedia content requires extra resources and specialized expertise, which may be beyond the scope of many studies focused on written or static visual methods. This type of content also may require more time and effort than textual or visual communication.
Secondly, evaluating the effectiveness of audio and video approaches poses methodological challenges, as it requires specific metrics to accurately assess user engagement and comprehension. Additionally, concerns regarding accessibility and inclusivity may discourage the adoption of multimedia strategies as researchers strive to accommodate diverse user needs. 
Finally, due to the lack of examples of multimedia communication in the security community, there may not exist enough empirical evidence demonstrating that multimedia communication is worth investing in over traditional methods in security contexts. As such, researchers continue to research more traditional communication methods. Future research should investigate the potential of audio and video communication to enhance security education and user engagement, addressing existing barriers to their adoption.

The limited use of audio and video communications suggests an opportunity to research \textbf{innovative communication methods}, especially in how they might cater to different learning styles or accessibility needs. However, some papers in our corpus also recommend establishing universal standards for security messages to help ensure clarity and consistency across platforms and technologies~\cite{wu2006security,habib2021toggles}. Railean et al.~\cite{railean2021onlite} suggest that governmental agencies should regulate and create uniform guidelines for security communication. We argue that balancing uniform guidelines while also allowing for innovation is a challenge for security communication.

\subsubsection{Empowering users}
Finally, just communicating about security is not enough if users cannot make informed choices because they do not have control over them. Future work should explore providing users with transparency through education about security choices and also \textbf{greater control}~\cite{weinshel2019oh,balebako2013little}. 

Security communication is only empowering if users have agency over their data and technology. The next step after educating users is building systems that allow users to express themselves. This is where usability comes into play. Future work should focus on how to design usable technology that enables users to express their preferences.

\section{Limitations}

Some factors may have influenced the comprehensiveness and generalizability of our findings. 
One significant limitation arises from the databases used for sourcing the papers. While we selected databases to provide a wide array of literature, they might have inherent biases. Similarly, despite our efforts, our search string could inadvertently exclude relevant studies due to the specificity or phrasing of the search terms. Such limitations could affect the breadth and depth of the gathered data. To mitigate this risk, we tried to include all the available databases and iterated heavily on the search string.

The inclusion criteria set for selecting the papers could further limit the study. We focus primarily on papers that explicitly discuss security communication. As such, we might have excluded valuable research that indirectly contributes to the field.  
Some studies we use to corroborate our findings have smaller sample sizes and a qualitative nature. So, although valuable for in-depth qualitative insights, they might not be generalizable to larger populations. Conversely, more extensive studies might prioritize breadth over depth, potentially overlooking detailed user interactions and nuanced behaviors.
Lastly, the studies in the review might contain biases based on their specific contexts, such as geographical location or user demographics. These biases could influence the applicability of findings across different contexts or cultures. However, we argue that we mitigate this risk by using several studies to corroborate our findings.

\section{Conclusion}

In this systematic literature review, we explored security communication. We attempted to provide a cohesive picture of the current state of security communication and listed seven actionable recommendations for future work. While traditional written methods dominate security communication, incorporating visual and interactive elements can enhance user engagement and understanding.  Our review also highlighted a preference for user-centered research designs, predominantly surveys and interviews, which may present an opportunity for future work to learn about security communication using other research methods. Moreover, we also found a significant reproducibility gap in existing studies, as few share their study protocols or artifacts, and almost no replication studies exist. Addressing this gap will be crucial for consolidating knowledge and improving the reliability of findings in future research.  Finally, we distilled advice for security communication from a corpus of 97 papers (identified from more than 3,400 candidate papers), such as the need for clear, but detailed language, the importance of context-specific information, and the benefits of personalizing the security information to the user's knowledge level and current needs.
We hope our work can be used to inform future research, strengthen security practices, empower users, and create a safer technological environment. 

\section{Acknowledgments}
This work was 
funded by Funda\c{c}\~ao para a Ci\^encia e a Tecnologia (FCT)
under grant PRT/BD/153739/2021,
and
projects UIDB/50021/2020 (DOI: 10.54499/UIDB/50021/2020), LA/P/0063/2020 (DOI: 10.54499/LA/P/0063/2020), the InfraGov project with reference 2024.07411.IACDC,  
and
the VeriFixer project, an FCT Exploratory Project with reference 2023.15557.PEX (DOI: 10.54499/2023.15557.PEX).

\bibliographystyle{ACM-Reference-Format}
\bibliography{bib}

\appendix

\section*{Full List of publications included in the review}\label{app:fulllist}

\begin{enumerate}
    \item Preliminary Insights in Security Warning Studies: An Exploration in University Context~\cite{emang2019preliminary}
    \item Appraisal on User's Comprehension in Security Warning Dialogs: Browsers Usability Perspective ~\cite{yi2020appraisal}
    \item Towards Improving the Efficacy of Windows Security Notifier for Apps from Unknown Publishers: The Role of Rhetoric~\cite{shrestha2023towards}
    \item Explanations in Warning Dialogs to Help Users Defend Against Phishing Attacks~\cite{desolda2023explanations}
    \item Stop Clicking on Update Later: Persuading Users, They Need Up-to-Date Antivirus Protection~\cite{zhang2014stopcomic}
    \item Enhancing the Usability of Android Application Permission Model~\cite{malik2019enhancing}
    \item Information Presentation: Considering On-line User Confidence for Effective Engagement~\cite{kani2015information}
    \item sD\&D: Design and Implementation of Cybersecurity Educational Game with Highly Extensible Functionality~\cite{kido2020sd}
    \item Android Permissions: User Attention, Comprehension, and Behavior~\cite{felt2012android}
    \item Replication: No One Can Hack My Mind Revisiting a Study on Expert and Non-Expert Security Practices and Advice~\cite{busse2019replication}
    \item Understanding Users' Knowledge about the Privacy and Security of Browser Extensions~\cite{kariryaa2021understanding}
    \item Effects of Explanatory Information on Privacy Policy Summarization Tool Perception~\cite{bracamonte2020effects}
    \item The Role of Instructional Design in Persuasion: A Comics Approach for Improving Cybersecurity~\cite{zhang2016rolecomics}
    \item Measuring the Effectiveness of Privacy Policies for Voice Assistant Applications~\cite{liao2020measuring}
    \item An Explainable Password Strength Meter Addon via Textual Pattern Recognition~\cite{xu2019explainable}
    \item Signalling Over-Privileged Mobile Applications Using Passive Security Indicators~\cite{iacono2017signalling}
    \item Design and Evaluation of a Data-Driven Password Meter~\cite{ur2017design}
    \item A Mobile App for Assisting Users to Make Informed Selections in Security Settings for Protecting Personal Health Data: Development and Feasibility Study~\cite{zhou2018mobile}
    \item Do Women in Conservative Societies (Not) Follow Smartphone Security Advice? A Case Study of Saudi Arabia and Pakistan~\cite{al2020women}
    \item ``It builds trust with the customers'' - Exploring User Perceptions of the Padlock Icon in Browser UI~\cite{von2022builds}
    \item ``What was that site doing with my Facebook password?'' Designing password-reuse notifications~\cite{golla2018site}
    \item ``Why Should I Read the Privacy Policy, I Just Need the Service'': A Study on Attitudes and Perceptions Toward Privacy Policies~\cite{ibdah2021should}
    \item ``I have no idea what they're trying to accomplish:'' Enthusiastic and Casual Signal Users' Understanding of Signal PINs~\cite{bailey2021have}
    \item ``I just looked for the solution!'' On Integrating Security-Relevant Information in Non-Security API Documentation to Support Secure Coding Practices~\cite{gorski2021just}
    \item A Brick Wall, a Locked Door, and a Bandit: A Physical Security Metaphor for Firewall Warnings~\cite{Raja_Hawkey_Hsu_Wang_Beznosov_2011}
    \item A Comprehensive Quality Evaluation of Security and Privacy Advice on the Web~\cite{frik2019privacy}
    \item A Look into User's Privacy Perceptions and Data Practices of IoT Devices~\cite{al2021look}
    \item A Study on Improving Security Warnings~\cite{zaaba2014study}
    \item A Systematic Evaluation of the Communicability of Online Privacy Mechanisms with Respect to Communication Privacy Management~\cite{coopamootoo2011systematic}
    \item An Experience Sampling Study of User Reactions to Browser Warnings in the Field~\cite{reeder2018experience}
    \item Anti-Phishing Phil: The Design and Evaluation of a Game That Teaches People Not to Fall for Phish~\cite{sheng2007anti}
    \item AppAware: A Policy Visualization Model for Mobile Applications~\cite{paspatis2020appaware}
    \item Browser Interfaces and Extended Validation SSL Certificates: An Empirical Study~\cite{biddle2009browser}
    \item Catering to Your Concerns: Automatic Generation of Personalised Security-Centric Descriptions for Android Apps~\cite{wu2019catering}
    \item Complex, but in a Good Way? How to Represent Encryption to Non-Experts Through Text and Visuals – Evidence from Expert Co-Creation and a Vignette Experiment~\cite{distler2022complex}
    \item Developers Deserve Security Warnings, Too: On the Effect of Integrated Security Advice on Cryptographic API Misuse~\cite{gorski2018developers}
    \item Developing Accessible and Usable Security (ACCUS) Heuristics~\cite{napoli2018developing}
    \item Developing Visualisations to Enhance an Insider Threat Product: A Case Study~\cite{graham2021developing}
    \item Do Security Toolbars Actually Prevent Phishing Attacks?~\cite{wu2006security}
    \item Does Context Influence Responses to Firewall Warnings?~\cite{mahmoud2012does}
    \item Does MoodyBoard Make Internet Use More Secure? Evaluating an Ambient Security Visualization Tool ~\cite{de2011does}
    \item A Nutrition Label for Privacy.~\cite{kelley2009nutrition}
    \item Effective Risk Communication for Android Apps~\cite{gates2013effective}
    \item Effects of the Design of Mobile Security Notifications and Mobile App Usability on Users' Security Perceptions and Continued Use Intention~\cite{wu2020effects}
    \item Enhancing Game-Based Learning Through Infographics in the Context of Smart Home Security~\cite{bahrini2020enhancing}
    \item Evaluating the End-User Experience of Private Browsing Mode~\cite{abu2020evaluating}
    \item Exploring User Reactions to New Browser Cues for Extended Validation Certificates~\cite{sobey2008exploring}
    \item Exploring User-Centered Security Design for Usable Authentication Ceremonies~\cite{fassl2021exploring}
    \item Faheem: Explaining URLs to People Using a Slack Bot~\cite{althobaiti2018faheem}
    \item From Secure to Military-Grade: Exploring the Effect of App Descriptions on User Perceptions of Secure Messaging~\cite{akgul2021secure}
    \item Gamification Techniques for Raising Cyber Security Awareness~\cite{scholefield2019gamification}
    \item HappyPermi: Presenting Critical Data Flows in Mobile Application to Raise User Security Awareness~\cite{bahrini2019happypermi}
    \item Help the User Recognize a Phishing Scam: Design of Explanation Messages in Warning Interfaces for Phishing Attacks~\cite{aneke2021help}
    \item How I Learned to be Secure: a Census-Representative Survey of Security Advice Sources and Behavior~\cite{redmiles2016learned}
    \item Human Relationships: A Never-Ending Security Education Challenge?~\cite{hagen2009human}
    \item I Think They're Trying to Tell Me Something: Advice Sources and Selection for Digital Security~\cite{redmiles2016think}
    \item Improving Computer Security Dialogs~\cite{bravo2011improving}
    \item Increasing Service Users' Privacy Awareness by Introducing On-Line Interactive Privacy Features ~\cite{kani2012increasing}
    \item Investigating Simple Privacy Indicators for Supporting Users When Installing New Mobile Apps ~\cite{stover2021investigating}
    \item Making Encryption Feel Secure: Investigating How Descriptions of Encryption Impact Perceived Security~\cite{distler2020making}
    \item Must I, Can I? I Don't Understand Your Ambiguous Password Rules~\cite{greene2017must}
    \item NetVisGame: Mobile Gamified Information Visualization of Home Network Traffic Data ~\cite{schufrin2022netvisgame}
    \item Oh, the Places You've Been! User Reactions to Longitudinal Transparency About Third-Party Web Tracking and Inferencing~\cite{weinshel2019oh}
    \item On the Comprehension of Security Risk Scenarios~\cite{hogganvik2005comprehension}
    \item OnLITE: On-Line Label for IoT Transparency Enhancement ~\cite{railean2021onlite}
    \item Personalized Systems Need Adaptable Privacy Statements!: How to Make Privacy-Related Legal Aspects Usable and Retraceable~\cite{garcia2009personalized}
    \item Privacy Pal: Improving Permission Safety Awareness of Third Party Applications in Online Social Networks~\cite{tucker2015privacy}
    \item PrivacyToon: Concept-driven Storytelling with Creativity Support for Privacy Concepts~\cite{suh2022privacytooninterative}
    \item Put Your Warning Where Your Link Is: Improving and Evaluating Email Phishing Warnings~\cite{petelka2019put}
    \item Revealing Hidden Context: Improving Mental Models of Personal Firewall Users~\cite{raja2009revealing}
    \item Smart Storytelling: Video and Text Risk Communication to Increase MFA Acceptability~\cite{das2020smart}
    \item Stopping Spyware at the Gate: A User Study of Privacy, Notice and Spyware~\cite{good2005stopping}
    \item Supporting Visual Security Cues for WebView-Based Android Apps~\cite{shin2013supporting}
    \item Symbolism in Computer Security Warnings: Signal Icons and Signal Words~\cite{samsudin2016symbolism}
    \item The Challenges of Understanding and Using Security: A Survey of End-Users~\cite{Furnell_Jusoh_Katsabas_2006}
    \item The Effect of Developer-Specified Explanations for Permission Requests on Smartphone User Behavior ~\cite{tan2014effect}
    \item The Effect of Entertainment Media on Mental Models of Computer Security~\cite{fulton2019effect}
    \item To Follow or Not to Follow: A Study of User Motivations Around Cybersecurity Advice~\cite{fagan2018follow}
    \item Toggles, Dollar Signs, and Triangles: How to(In)Effectively Convey Privacy Choices with Icons and Link Texts~\cite{habib2021toggles}
    \item Towards Automatic Generation of Security-Centric Descriptions for Android Apps~\cite{zhang2015towards}
    \item Towards XAI in the SOC – A User Centric Study of Explainable Alerts with SHAP and LIME~\cite{eriksson2022towards}
    \item Understanding Perceptions: User Responses to Browser Warning Messages ~\cite{molyneaux2019understanding}
    \item Understanding Sensor Notifications on Mobile Devices~\cite{ma2017understanding}
    \item Usable Security: A Browser’s Security Warnings Assessment~\cite{emang2020usable}
    \item Usable Security: Revealing End-Users Comprehensions on Security Warnings~\cite{amran2017usable}
    \item User-Centered Design of Visualizations for Software Vulnerability Reports~\cite{reynolds2021user}
    \item Using Cartoons to Teach Internet Security~\cite{srikwan2008using}
    \item Using Statistical Information to Communicate Android Permission Risks to Users~\cite{kraus2014using}
    \item Visual Analytics and Visualization for Android Security Risk~\cite{yoo2019visual}
    \item What Could Go Wrong? Raising Mobile Privacy and Security Awareness Through a Decision-Making Game~\cite{zargham2019could}
    \item What Risk? I Don't Understand. An Empirical Study on Users' Understanding of the Terms Used in Security Texts~\cite{wu2020risk}
    \item Why Do They Do What They Do? A Study of What Motivates Users to (Not) Follow Computer Security Advice~\cite{fagan2016they}
    \item Will You Trust This TLS Certificate? Perceptions of People Working in IT~\cite{ukrop2020will}
    \item Designing a Privacy Label: Assisting Consumer Understanding of Online Privacy Practices~\cite{kelley2009designing}
    \item Little Brothers Watching You: Raising Awareness of Data Leaks on Smartphones~\cite{balebako2013little}
    \item Privacy Explanations—A Means to End-User Trust~\cite{brunotte2023privacy}
    \item Your Secrets Are Safe: How Browsers' Explanations Impact Misconceptions About Private Browsing Mode~\cite{wu2018your}
    
\end{enumerate}

\end{document}